\documentclass[11pt]{article}

\usepackage[utf8]{inputenc}
\usepackage[T1]{fontenc}
\usepackage[margin=1in]{geometry}
\usepackage{amsmath,amssymb}
\usepackage{booktabs}
\usepackage{graphicx}
\usepackage{xcolor}
\usepackage{caption}
\usepackage[hidelinks]{hyperref}
\usepackage{orcidlink}

\graphicspath{{figures/}}

\newcommand{\alch}{a_{\mathrm{lch}}}
\newcommand{\blch}{b_{\mathrm{lch}}}
\newcommand{\alab}{a_{\mathrm{lab}}}
\newcommand{\blab}{b_{\mathrm{lab}}}

\title{Chroma-gated, differentiable OKLCH interpolation:\\
Continuous Oklab fallback for color-cast reduction%
\thanks{Licensed under CC BY 4.0.}}
\author{Naoyuki Uchida\\
  \small Independent Researcher, Japan\\
  \small \orcidlink{0000-0002-9551-0470}~\href{https://orcid.org/0000-0002-9551-0470}{0000-0002-9551-0470}\quad
  \href{mailto:naoyuki.uchida.DNF@gmail.com}{naoyuki.uchida.DNF@gmail.com}}
\date{2026}

\begin{document}
\maketitle

\begin{abstract}
OKLCH---the cylindrical (lightness, chroma, hue) form of Ottosson's Oklab color
space---is the interpolation space recommended by CSS Color~4 for gradients and
\texttt{color-mix()}, and it is now broadly deployed. Its polar parameterization,
however, casts color near the neutral axis in two ways: an \emph{(1) inter-hue detour}
between two chromatic endpoints that sweeps through an unintended hue (blue\,$\to$\,yellow
visibly passing through green), and an \emph{(2) off-line bow} when one endpoint is
achromatic. Existing remedies are uniformly \emph{two-valued}---a threshold switch that
fires only at an achromatic endpoint---so they address only~(2); on chromatic pairs
every one of them reduces to raw OKLCH, leaving the (1) inter-hue cast untreated. We
introduce \textbf{Continuous Oklab fallback (COFb)}, a one-parameter, differentiable
chroma gate $w(C)=C^{n}/(C^{n}+\sigma^{n})$ that continuously blends the OKLCH path
toward the linear Oklab path as chroma falls. A single gate reduces the (1) cast that
the two-valued family leaves untreated and unifies the handling of (1) and~(2) without
any endpoint test. We characterize a cast--hue trade-off frontier, adopt a default
($n=1$, the rational Michaelis--Menten form; $\sigma\approx0.19$ for a typical sRGB
palette, from a normalization-independent cast-half criterion), and verify the gate's
properties symbolically. At the default, COFb halves the inter-hue path detour (mean lateral
deviation $-49.5\%$, chroma-weighted hue excursion $-35.5\%$). We also state the
method's limits: on~(2) alone the two-valued switch remains better, and like
any Cartesian blend COFb does not preserve chroma. In deployment, COFb runs entirely in plain Oklab $(a,b)\to$ sRGB, so it serves as a fallback that delivers the same cast-reduced gradients where modern CSS color interpolation (\texttt{color-mix(in oklch)} and the like) is unavailable---older engines, image and video pipelines, or GPU shaders.
\end{abstract}

\section{Introduction}
\label{sec:intro}

Perceptual color spaces have become the default substrate for color interpolation on
the web. OKLCH---the cylindrical (lightness, chroma, hue) form of Ottosson's
Oklab~\cite{ottosson2020}---is the interpolation space recommended by CSS Color~4 for
gradients and \texttt{color-mix()}~\cite{csscolor4}, and it is now widely deployed in
practice (for example, as the default palette space in recent CSS tooling such as
Tailwind~v4~\cite{adoption}). Its appeal is straightforward: interpolating along hue at
constant chroma traces the intuitive arc a designer expects, rather than the
desaturated straight line a Cartesian blend would produce.

That same polar parameterization, however, carries a well-known pathology near the
neutral axis. At zero chroma the hue angle is undefined, and in its neighborhood a
small change in the $(a,b)$ coordinates produces an arbitrarily large change in hue.
Two visible artifacts follow from this single geometric fact, and it is useful to name
them separately:

\begin{itemize}
  \item \textbf{(1) Inter-hue color cast.} When both endpoints are chromatic but lie
  on opposite sides of the neutral axis (for example, blue\,$\to$\,yellow), the
  constant-radius hue arc detours \emph{around} the origin. The interpolated path
  bulges in chroma and sweeps through an unintended hue---blue\,$\to$\,yellow visibly
  passes through green. Both endpoints have a perfectly well-defined hue; the artifact
  lives in the path between them.
  \item \textbf{(2) Achromatic-endpoint cast.} When one endpoint is achromatic (for
  example, green\,$\to$\,black), its hue is genuinely undefined, yet raw OKLCH still
  carries a hue value forward, swinging the mid-path away from the straight
  chromatic-to-neutral line.
\end{itemize}

Existing remedies target~(2) and are uniformly \emph{two-valued}: they apply a
threshold test, and when an endpoint is judged achromatic they switch that endpoint's
hue to a defined value. The CSS \emph{missing-component} mechanism treats an undefined
hue as \texttt{none} (NaN) and inherits it from the other endpoint during
interpolation~\cite{csscolor4,csswg6107}; ColorAide's \texttt{is\_achromatic()}
threshold likewise marks the hue \emph{powerless} and carries it forward~\cite{coloraide}.
The shared intuition---that chroma should down-weight the influence of hue---is correct
in spirit. But its realization is a hard switch that fires only at an achromatic
endpoint. Against the (1) inter-hue cast it does nothing at all: both endpoints are
chromatic, the threshold never trips, and every two-valued method reduces exactly to
raw OKLCH there. The (1) cast is a blind spot of the entire existing family.

We close this gap with \textbf{Continuous Oklab fallback (COFb)}: a one-parameter,
differentiable chroma gate that continuously blends the raw OKLCH path toward the
linear Oklab path as chroma falls. The gate $w(C)=C^{n}/(C^{n}+\sigma^{n})$ weights the
OKLCH path ($w\to1$ at high chroma, full hue control) against linear Oklab ($w\to0$ at
the neutral axis, no cast), and the blend is taken in the Cartesian $(a,b)$ plane.
Because the gate responds to the \emph{interpolated} chroma at every point along the
path, it acts wherever chroma is low---near an achromatic endpoint~(2) and along the
origin-skirting detour of an inter-hue arc~(1) alike---without any endpoint test.

\paragraph{Contributions.}
\begin{itemize}
  \item \textbf{(C1)} COFb: a continuous, differentiable, chroma-gated interpolation
  that spans raw OKLCH ($\sigma\to0$) and linear Oklab ($\sigma\to\text{large}$) as a
  single family. The gate's properties are confirmed symbolically
  (Section~\ref{sec:method}).
  \item \textbf{(C2)} COFb is the \emph{only} method among those compared that
  addresses the (1) inter-hue cast; the existing two-valued remedies reduce to raw
  OKLCH on chromatic pairs. At the recommended default it roughly halves the inter-hue
  path detour (mean lateral deviation $-49.5\%$, chroma-weighted hue excursion
  $-35.5\%$; Section~\ref{sec:eval}).
  \item \textbf{(C3)} A single continuous gate \emph{unifies (1) and~(2)}, replacing
  the special-cased, two-valued, (2)-only handling with one expression
  (Section~\ref{sec:eval}).
  \item \textbf{(C4)} A parameter study exposes a cast--hue \emph{trade-off frontier}:
  removing more cast necessarily relaxes hue control. A normalization-independent
  cast-reduction criterion fixes an interpolation default of $\sigma\approx0.19$, and
  the optimum is practically insensitive to~$n$ (it moves only at the error level between $n=1$ and $0.87$); we adopt $n=1$ (the Michaelis--Menten form
  $w=C/(C+\sigma)$), which is rational, GPU-friendly, and $C^{1}$ up to the endpoint
  (Section~\ref{sec:eval}).
\end{itemize}

\paragraph{Scope and claims.} COFb is a \emph{geometric, behavior-level}
construction; we make no claim that its paths agree with a true perceptual metric. Two
limitations are stated explicitly and revisited in Section~\ref{sec:discussion}. First, on
the (2) achromatic case the two-valued methods drive the lateral deviation to exactly
zero, which COFb improves upon but does not match---(2), taken alone, is still better
served by the existing switch. Second, like any Cartesian blend, COFb does not preserve
chroma: the path grays slightly as it nears the neutral axis. The contribution is a
continuous remedy for the cast that the two-valued family leaves untreated, not a claim
of perceptual optimality.

\section{Background and Related Work}
\label{sec:background}

\subsection{Oklab and OKLCH}
Oklab is a perceptual color space designed by Ottosson for image-processing and
interpolation use, fitting a lightness/opponent representation to perceptual data so
that Euclidean operations behave reasonably~\cite{ottosson2020}. OKLCH is its
cylindrical form, replacing the opponent pair $(a,b)$ with chroma and hue,
$C=\sqrt{a^{2}+b^{2}}$ and $h=\operatorname{atan2}(b,a)$. CSS Color~4 adopts these
spaces and recommends OKLCH as an interpolation space for gradients and
\texttt{color-mix()}~\cite{csscolor4}. The motivation for interpolating in OKLCH rather
than Oklab is hue control: a constant-chroma sweep in $h$ follows the arc a designer
expects, whereas a straight $(a,b)$ segment cuts across it and desaturates. OKLCH is
now broadly deployed---supported across major browsers and adopted as a default palette
space in widely used CSS tooling~\cite{adoption}---so the setting in which this
interpolation runs is a common one.

\subsection{The achromatic hue singularity}
OKLCH's cylindrical parameterization, like any polar parameterization, is singular on the neutral axis. At $C=0$ the hue
$h=\operatorname{atan2}(b,a)$ is undefined, and in any neighborhood of the axis a small
displacement in $(a,b)$ produces an arbitrarily large change in $h$. This single fact
produces two distinct interpolation artifacts (the naming used throughout this paper):
\emph{(1) inter-hue cast}---both endpoints chromatic but on opposite sides of the axis;
the constant-radius arc detours around the origin, bulging in chroma and sweeping
through an unintended hue (e.g.\ blue\,$\to$\,yellow through green); and
\emph{(2) achromatic-endpoint cast}---one endpoint achromatic; its hue is genuinely
undefined, yet a value is still carried forward, bending the path off the
chromatic-to-neutral line.

\subsection{Two-valued handling}
Existing remedies all target~(2) and share one mechanism: a threshold test that, when
an endpoint is judged achromatic, replaces its undefined hue with a defined one.

\textbf{CSS missing components.} This handling originates in CSS Working Group issue
\#6107, which proposed treating the undefined hue of an achromatic color as
\emph{missing} rather than as NaN; the issue is closed (resolved by editor
discretion)~\cite{csswg6107}. The specified behavior is that, during interpolation, a
missing hue at one endpoint is inherited from the other, and if both are missing the
hue is taken as zero~\cite{csscolor4}. A follow-up question on whether achromatic
interpolation should take the longer hue arc---issue \#9436, extending the earlier
\#9224 discussion of transparent colors---was likewise closed, confirming the existing
undefined-hue interpolation logic~\cite{csswg9436}. All of this concerns the (2)
achromatic-endpoint case; the (1) inter-hue cast is not discussed.

\textbf{ColorAide.} The library's \texttt{is\_achromatic()} test uses a per-space
chroma threshold to detect colors at or very near the neutral axis and sets their hue
undefined; it then offers CSS-compatible \emph{powerless} and \emph{carryforward}
policies that drop or inherit the undefined hue during interpolation~\cite{coloraide}.
Its stated rationale---that as chroma approaches zero the hue matters less, so dropping
it near the axis avoids odd color shifts---is the same intuition that motivates COFb.

The shared intuition (chroma should down-weight the influence of hue) is therefore
correct in spirit, and COFb can be read as its continuous realization
(Section~\ref{sec:discussion}). In every existing case, however, the realization is a
\emph{two-valued switch}---hue is either defined or replaced, with nothing in
between---and the switch fires \emph{only when an endpoint is achromatic}.

\subsection{The gap}
Because the two-valued switch is conditioned on an achromatic \emph{endpoint}, it never
engages on the (1) inter-hue case, where both endpoints are chromatic and the cast
lives in the path between them. On chromatic pairs every two-valued method evaluates
identically to raw OKLCH---a fact we confirm numerically in Section~\ref{sec:eval}
(Table~\ref{tab:metrics}). Two gaps therefore remain in the existing family: (i) there
is no continuous treatment, only a hard switch; and (ii) the (1) inter-hue cast is left
entirely untreated. COFb addresses both with a single chroma-gated, differentiable
blend (Section~\ref{sec:method}). Since the relevant CSS issues are closed, we frame
COFb as a \emph{continuous version of missing-component handling} at the implementation
level.

\section{Method: Continuous Oklab fallback}
\label{sec:method}

\subsection{Construction}
Let the two endpoints be given in OKLCH and let $t\in[0,1]$ parameterize the
interpolation. We compute two reference paths in Oklab coordinates: the \emph{raw
OKLCH path} $(\alch,\blch)$, obtained by interpolating lightness and chroma linearly
and hue along the shorter arc, then converting $(L,C,h)\to(a,b)$; and the \emph{linear
Oklab path} $(\alab,\blab)$, a straight segment between the two endpoints in the
$(a,b)$ plane. Lightness is interpolated linearly in both and is shared. COFb blends
the two chromatic paths point-by-point with a chroma-dependent gate $w$:
\begin{equation}
a(t)=w\,\alch(t)+(1-w)\,\alab(t),\quad
b(t)=w\,\blch(t)+(1-w)\,\blab(t),\quad
L(t)=(1-t)L_{0}+t\,L_{1}.
\end{equation}
The gate is the Naka--Rushton / Michaelis--Menten saturation function of the chroma
along the raw OKLCH path, $C(t)=\sqrt{\alch^{2}+\blch^{2}}$:
\begin{equation}
w(C)=\frac{C^{n}}{C^{n}+\sigma^{n}},\qquad \sigma>0,\; n>0.
\label{eq:gate}
\end{equation}
At high chroma $w\to1$ and the path follows OKLCH (full hue control); as chroma falls
toward the neutral axis $w\to0$ and the path relaxes toward the cast-free linear Oklab
segment. Because the gate reads the \emph{interpolated} chroma at each $t$, it engages
wherever chroma is low---at an achromatic endpoint~(2) and along the origin-skirting
detour of an inter-hue arc~(1) alike---with no endpoint test and no branching.

\subsection{Properties (symbolically verified)}
For $C>0,\,n>0,\,\sigma>0$ the gate~\eqref{eq:gate} satisfies the properties in
Table~\ref{tab:props}. All were confirmed with a computer-algebra system (SymPy); the
verification log is reproduced in Appendix~\ref{app:repro}.

\begin{table}[ht]
\centering
\caption{Properties of the chroma gate~\eqref{eq:gate} for $C>0$.}
\label{tab:props}
\begin{tabular}{ll}
\toprule
property & result \\
\midrule
endpoint values & $w(0^{+})=0,\; w(\sigma)=\tfrac12,\; w(C\to\infty)=1$ \\
monotonicity & $w'(C)=\dfrac{n\,C^{n-1}\sigma^{n}}{(C^{n}+\sigma^{n})^{2}}>0$ (strictly increasing) \\
smoothness on $C>0$ & analytic (all derivatives exist and are finite) \\
\bottomrule
\end{tabular}
\end{table}

The parameter $\sigma$ is the chroma at which the gate is half open, giving it a direct
operational meaning. The $C=0$ endpoint behavior is \emph{$n$-dependent}, and this is
the property that selects the default:
\begin{equation}
\lim_{C\to0^{+}}w'(C)=
\begin{cases}
+\infty & n<1 \quad(\text{vertical tangent; }C^{1}\text{ fails at }C=0),\\[2pt]
1/\sigma & n=1 \quad(\text{finite slope; }C^{1}\text{ up to the endpoint}),\\[2pt]
0 & n>1.
\end{cases}
\end{equation}
For $n<1$ the gate is continuous at $C=0$ ($w(0)=0$) but not $C^{1}$: its one-sided
slope is infinite. This is not a defect---it is the same vertical tangent at $C=0$ (a
steep near-origin response) used for the C-axis compression in the author's
Oklch+~\cite{oklchplus}---but it does mean a
careful statement is required: COFb is differentiable on $C>0$ for every $n$, and
additionally $C^{1}$ at the neutral endpoint only for $n\ge1$.

\subsection{Default: $n=1$ (Michaelis--Menten)}
\label{sec:default}
We adopt $n=1$, for which the gate reduces to the Michaelis--Menten form
\begin{equation}
w(C)=\frac{C}{C+\sigma},\qquad w'(C)=\frac{\sigma}{(C+\sigma)^{2}},\qquad
w'(0^{+})=\frac{1}{\sigma}.
\label{eq:mm}
\end{equation}
Three reasons. (i) \emph{Endpoint regularity}: $w'(0^{+})=1/\sigma$ is finite, so the
gate is $C^{1}$ all the way to the neutral axis, removing the vertical tangent present
for $n<1$. (ii) \emph{Rational, GPU-friendly form}: $C/(C+\sigma)$ contains no
fractional powers; the forward gate and its closed-form inverse $C=\sigma\,w/(1-w)$ are
both rational. (For general $n$ the inverse is $C=\sigma\,(w/(1-w))^{1/n}$, which loses
rationality.) (iii) \emph{Consistency}: it is the same Michaelis--Menten functional
form used for the C-axis compression in the author's Oklch+~\cite{oklchplus}, keeping the toolset coherent.

The exponent $n=0.87$ is retained as a documented alternative. It is the C-axis
Naka--Rushton compression fit to COMBVD color-difference data in the author's
Oklch+~\cite{oklchplus} (as is the reused $\sigma=0.34$); here we reuse only the
\emph{functional form} as an interpolation gate, not the C-axis compression itself. It
has a
vertical tangent at $C=0$, and---crucially---the optimum of
$\sigma$ is \emph{practically insensitive to $n$} over the tested range (Section~\ref{sec:paramstudy}), so the
interpolation behavior of the two choices differs only at the error level. We therefore
default to the structurally cleaner $n=1$ and note $n=0.87$ as a near-equivalent
option.

\subsection{Limiting behavior}
COFb is a continuous one-parameter family bridging the two standard choices:
\[
\sigma\to0\;\Rightarrow\;w\equiv1\;\Rightarrow\;\text{raw OKLCH (maximum hue control, full cast)};
\]
\[
\sigma\to\text{large}\;\Rightarrow\;w\equiv0\;\Rightarrow\;\text{linear Oklab (no cast, no hue control)}.
\]
The interpolation default $\sigma\approx0.19$ (Section~\ref{sec:paramstudy}) sits
between these extremes, at the chroma-half point that halves the inter-hue cast. Unlike
the existing two-valued fallback---which is a discrete switch between exactly these two
endpoints, triggered only by an achromatic endpoint---COFb traverses the family
continuously and as a function of the path's own chroma.

\section{Evaluation}
\label{sec:eval}

\subsection{Metrics}
We measure interpolation geometry directly in the Oklab $(a,b)$ plane---the coordinate
system in which the blend takes place---so the metrics are independent of any
color-difference formula such as CIEDE2000. Let $\mathbf{p}(t)=(a(t),b(t))$ be a path in
the $(a,b)$ plane and $\mathbf{p}_{\mathrm{lin}}(t)$ the straight linear-Oklab reference
between the same endpoints $P_{0}=\mathbf{p}(0)$, $P_{1}=\mathbf{p}(1)$ (the cast-free
reference), with $C(t)=\lVert\mathbf{p}(t)\rVert$ and
$h(t)=\operatorname{atan2}(b(t),a(t))$. Paths are sampled at $N=257$ equal-$t$ points and
all maxima and means are taken over the samples.

Our \emph{primary metric} is the \emph{lateral deviation}, the maximum perpendicular
distance from the chord joining the two endpoints:
\begin{equation}
D_{\mathrm{lat}}=\max_{t}\bigl\lVert(\mathbf{p}(t)-P_{0})-\langle\mathbf{p}(t)-P_{0},\hat{\mathbf{d}}\rangle\,\hat{\mathbf{d}}\bigr\rVert,
\qquad \hat{\mathbf{d}}=\frac{P_{1}-P_{0}}{\lVert P_{1}-P_{0}\rVert}.
\label{eq:lateral}
\end{equation}
It captures both artifacts---the origin detour of~(1) and the off-line bow of~(2).

We decompose this deviation into two mechanistic components, measured against the same
reference at equal $t$: \emph{excess chroma} (the radial bulge of the detour) and
\emph{hue excursion} (the chroma-weighted mean angular deviation of the path hue,
``blue\,$\to$\,yellow passing through green''):
\begin{equation}
\Delta C_{\max}=\max_{t}\bigl(C(t)-C_{\mathrm{lin}}(t)\bigr),\qquad
H=\frac{\sum_{t}\lvert\Delta h(t)\rvert\,C(t)}{\sum_{t}C(t)},
\label{eq:decomp}
\end{equation}
with the shortest-arc hue difference
$\Delta h(t)=\bigl((h(t)-h_{\mathrm{lin}}(t)+180)\bmod 360\bigr)-180$, where
$C_{\mathrm{lin}},h_{\mathrm{lin}}$ are the chroma and hue of the linear reference
and the weight is the path's own chroma $C(t)$, so that points near the achromatic
axis---where hue is ill-defined and barely visible---contribute little. These two are
not independent of $D_{\mathrm{lat}}$: they read the same deviation along its radial and
angular directions, and Section~\ref{sec:results} shows how they partition the two pair
groups.

\subsection{Methods}
We compare, over a fixed set of endpoint pairs: (0)~\emph{linear Oklab}, a straight
$(a,b)$ segment (cast-free reference, no hue control); (i)~\emph{raw OKLCH}, polar
interpolation (the artifact under study); (ii)~\emph{NaN-inherit}, CSS
missing-component handling (achromatic hue inherited); (iii)~\emph{two-valued
fallback}, linear Oklab if an endpoint is achromatic, else raw OKLCH; and
(iv)~\emph{COFb}, this work, at the default $n=1,\,\sigma=0.19$ ($n=0.87$ reported as an
alternative). Pairs are split into \emph{(1) inter-hue}
(Blue\,$\to$\,Yellow, Cyan\,$\to$\,Red, Purple\,$\to$\,Orange) and
\emph{(2) achromatic-endpoint} (Green\,$\to$\,Black, Blue\,$\to$\,Black). The OKLCH
endpoint coordinates are given in Table~\ref{tab:pairs}, so Table~\ref{tab:metrics} is
reproducible from the text alone.

\begin{table}[ht]
\centering
\caption{OKLCH endpoint coordinates $(L,C,h)$ ($h$ in degrees). Achromatic endpoints
have $C=0$ with $h$ undefined.}
\label{tab:pairs}
\small
\begin{tabular}{lll}
\toprule
group, pair & endpoint 0 $(L,C,h)$ & endpoint 1 $(L,C,h)$ \\
\midrule
(1) Blue\,$\to$\,Yellow   & $(0.45,\,0.22,\,264)$ & $(0.92,\,0.19,\,100)$ \\
(1) Cyan\,$\to$\,Red      & $(0.78,\,0.14,\,195)$ & $(0.58,\,0.22,\,28)$ \\
(1) Purple\,$\to$\,Orange & $(0.50,\,0.20,\,310)$ & $(0.72,\,0.18,\,55)$ \\
(2) Green\,$\to$\,Black   & $(0.60,\,0.16,\,145)$ & $(0.00,\,0.00,\,\text{n/a})$ \\
(2) Blue\,$\to$\,Black    & $(0.45,\,0.22,\,264)$ & $(0.00,\,0.00,\,\text{n/a})$ \\
\bottomrule
\end{tabular}
\end{table}

\subsection{Results}
\label{sec:results}
Table~\ref{tab:metrics} reports all three metrics for the five methods on both pair
groups, and Table~\ref{tab:groupmean} the group-mean lateral deviation and hue excursion.

\begin{table}[ht]
\centering
\caption{Interpolation metrics (COFb at the default: $n=1,\,\sigma=0.19$). Lateral deviation
and excess chroma are in Oklab units; hue excursion in degrees.}
\label{tab:metrics}
\small
\begin{tabular}{lllrrr}
\toprule
group & pair & method & lateral dev. & excess $C$ & hue exc.\ ($^{\circ}$) \\
\midrule
(1) & Blue\,$\to$\,Yellow & Oklab-linear & 0.0000 & 0.0000 & 0.00 \\
    &                     & Oklch-raw    & 0.1768 & 0.1756 & 25.42 \\
    &                     & NaN-inherit  & 0.1768 & 0.1756 & 25.42 \\
    &                     & Two-valued   & 0.1768 & 0.1756 & 25.42 \\
    &                     & \textbf{COFb} & \textbf{0.0919} & \textbf{0.0908} & \textbf{17.02} \\
(1) & Cyan\,$\to$\,Red    & Oklch-raw    & 0.1620 & 0.1521 & 30.86 \\
    &                     & COFb         & 0.0798 & 0.0714 & 19.60 \\
(1) & Purple\,$\to$\,Orange & Oklch-raw  & 0.0746 & 0.0742 & 4.72 \\
    &                     & COFb         & 0.0373 & 0.0370 & 2.70 \\
(2) & Green\,$\to$\,Black & Oklch-raw    & 0.0814 & 0.0000 & 48.14 \\
    &                     & NaN-inh.\,/\,Two-val. & 0.0000 & 0.0000 & 0.00 \\
    &                     & COFb         & 0.0287 & 0.0000 & 11.93 \\
(2) & Blue\,$\to$\,Black  & Oklch-raw    & 0.0828 & 0.0000 & 31.87 \\
    &                     & NaN-inh.\,/\,Two-val. & 0.0000 & 0.0000 & 0.00 \\
    &                     & COFb         & 0.0340 & 0.0000 & 10.57 \\
\bottomrule
\end{tabular}
\end{table}

\begin{table}[ht]
\centering
\caption{Group-mean lateral deviation and hue excursion (lateral deviation in Oklab
units; hue excursion in degrees).}
\label{tab:groupmean}
\begin{tabular}{lrrrr}
\toprule
 & \multicolumn{2}{c}{(1) inter-hue} & \multicolumn{2}{c}{(2) achromatic} \\
\cmidrule(lr){2-3}\cmidrule(lr){4-5}
method & lat.\ dev. & hue exc.\ ($^{\circ}$) & lat.\ dev. & hue exc.\ ($^{\circ}$) \\
\midrule
Oklch-raw & 0.1378 & 20.3 & 0.0821 & 40.0 \\
NaN-inherit / Two-valued & 0.1378 & 20.3 & 0.0000 & 0.0 \\
\textbf{COFb} ($n=1$) & \textbf{0.0697} & \textbf{13.1} & 0.0313 & 11.2 \\
COFb ($n=0.87$, alt) & 0.0695 & 13.1 & 0.0324 & 12.0 \\
\bottomrule
\end{tabular}
\end{table}

\begin{figure}[ht]
\centering
\includegraphics[width=\linewidth]{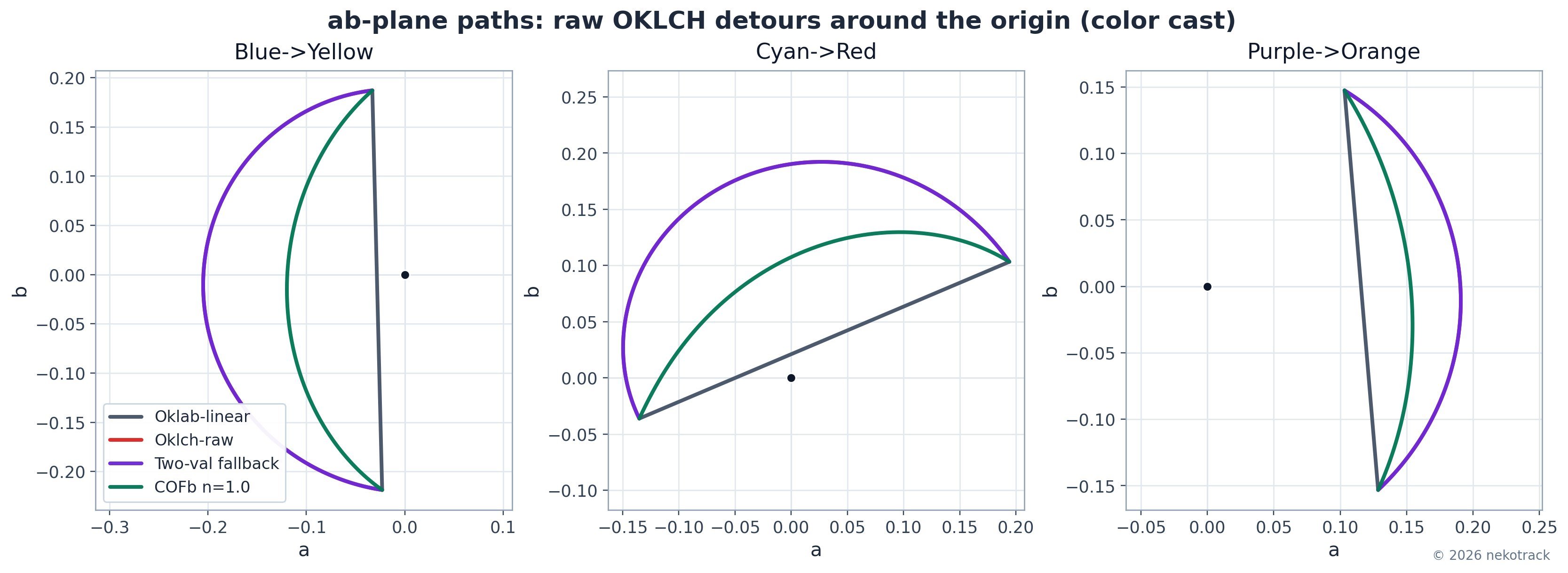}
\caption{Interpolation paths in the Oklab $(a,b)$ plane. Raw OKLCH (and the two-valued
methods, identical on chromatic pairs) detours around the origin; COFb relaxes the
detour toward the cast-free linear segment.}
\label{fig:abpaths}
\end{figure}

\begin{figure}[ht]
\centering
\includegraphics[width=\linewidth]{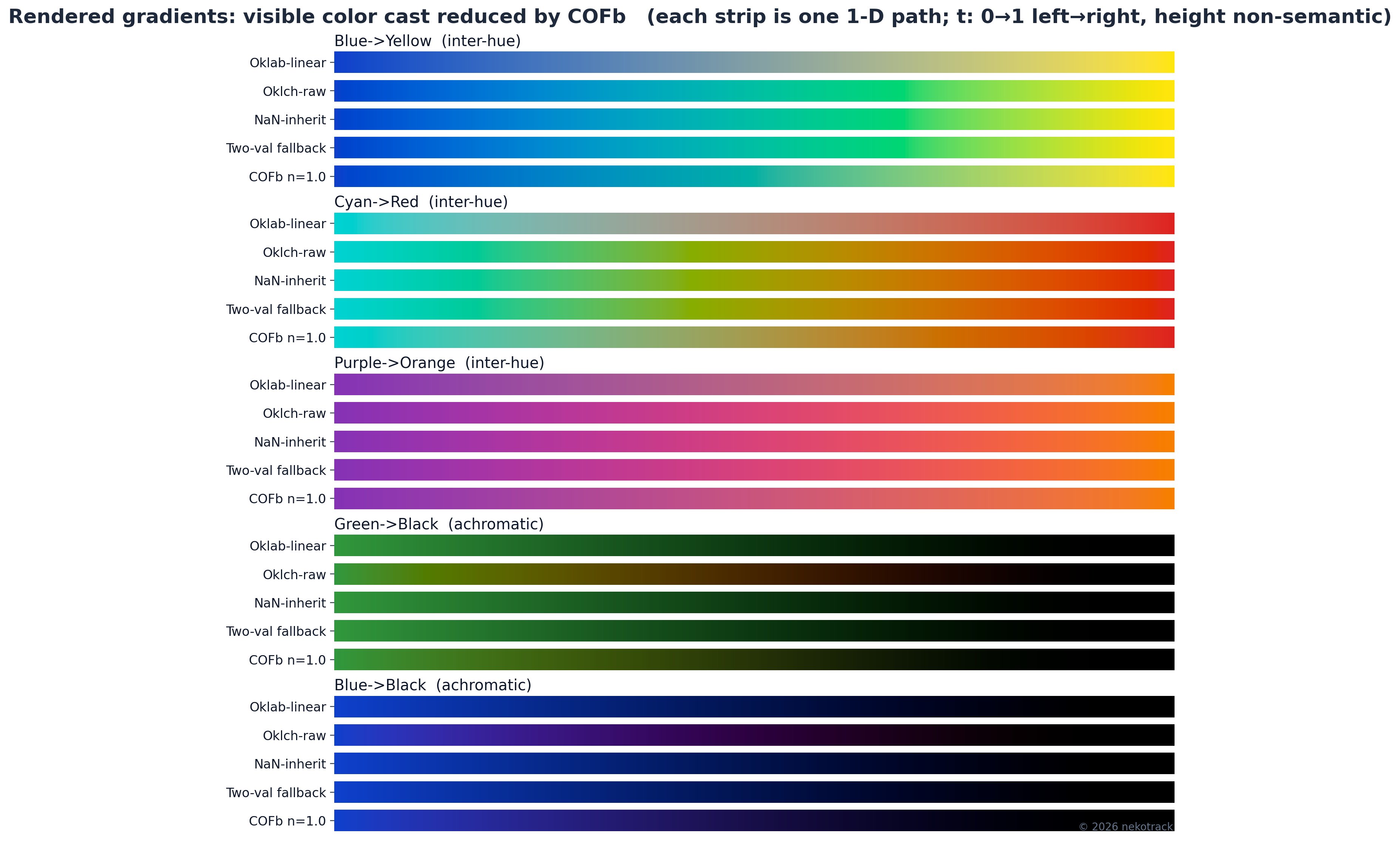}
\caption{Rendered gradients (sRGB). For the inter-hue pairs, raw OKLCH, NaN-inherit, and
the two-valued fallback all sweep through an unintended mid-path hue; COFb visibly reduces the cast. For the achromatic pairs, COFb stays close to the chromatic-to-neutral
line.}
\label{fig:strips}
\end{figure}

\paragraph{(1) Inter-hue.} On chromatic pairs the two-valued methods are numerically
identical to raw OKLCH---their achromatic test never fires---so every existing method
shares raw OKLCH's full cast. COFb is the only method that acts here, reducing the mean
lateral deviation by $49.5\%$ ($0.1378\to0.0697$) and the chroma-weighted hue excursion
by $35.5\%$. (By construction $\sigma=0.19$ is the cast-half point, so the $\sim$50\%
reduction and the $\sigma$ criterion of Section~\ref{sec:paramstudy} are mutually
consistent.) Across these pairs excess chroma and lateral deviation nearly coincide
(e.g.\ $0.1756$ vs $0.1768$ for Blue$\to$Yellow), confirming that the detour is
essentially a pure radial bulge. The rendered gradient strips show the corresponding visible cast reduced
(Figures~\ref{fig:abpaths} and~\ref{fig:strips}).

\paragraph{(2) Achromatic endpoint.} Here NaN-inherit and the two-valued fallback drive
the lateral deviation to exactly~0 (full linear Oklab); COFb improves substantially
over raw OKLCH ($0.0821\to0.0313$) but does not reach~0 (Appendix
Fig.~\ref{fig:achropaths}). We note this explicitly: taken
alone, the achromatic case is better served by the existing two-valued handling. Excess
chroma is identically zero for these pairs---a chromatic-to-achromatic path stays within
the chroma envelope---so the residual artifact is the off-line bow together with the
carried-over hue, not a radial bulge. The
contribution of COFb is that one continuous gate covers both groups without a switch
(Sections~\ref{sec:crosscheck} and~\ref{sec:discussion}). The $n=0.87$ alternative
differs from $n=1$ only at the error level on every pair, supporting $n=1$ as the
default.

\subsection{Parameter study: the cast--hue trade-off}
\label{sec:paramstudy}
Sweeping $\sigma$ traces a trade-off frontier: smaller $\sigma$ keeps more of OKLCH's
hue control but removes less cast; larger $\sigma$ removes more cast but flattens toward
linear Oklab. The ideal upper-left corner (full cast removal \emph{and} full hue
control) is unreachable. Table~\ref{tab:frontier} summarizes two operating points (cast
here is the mean inter-hue lateral deviation; raw baseline $=0.1378$).

\begin{table}[ht]
\centering
\caption{Operating points on the cast--hue frontier ($n=1$).}
\label{tab:frontier}
\begin{tabular}{lrrrr}
\toprule
criterion & $\sigma^{*}$ & inter-hue cast & hue ret.\ (gate) & hue ret.\ (path) \\
\midrule
(A) hue-priority: gate retention $=0.5$ & 0.144 & 0.079 & 0.50 & 0.72 \\
(B) cast-half: cast $=0.5\times$ baseline & 0.194 & 0.069 & 0.43 & 0.66 \\
\bottomrule
\end{tabular}
\end{table}

Here hue retention (gate) is the mean gate $w$ over the representative chromas
$C\in\{0.10,\,0.15,\,0.20\}$, and hue retention (path) is the path-based quantity defined
in Section~\ref{sec:crosscheck}.

\begin{figure}[ht]
\centering
\includegraphics[width=0.86\linewidth]{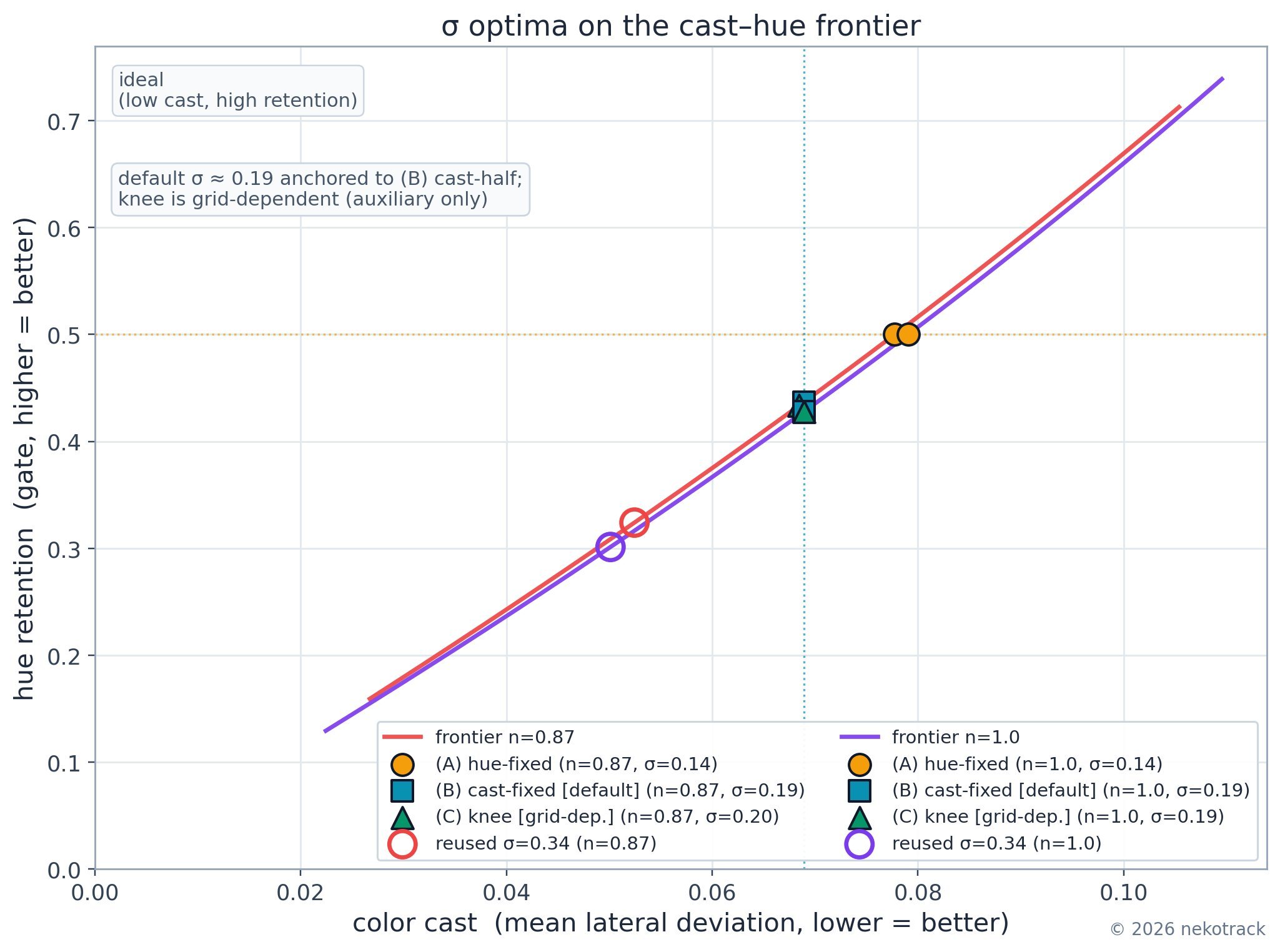}
\caption{The cast--hue frontier with operating points. The default $\sigma\approx0.19$
is anchored to criterion~(B) (cast-half); the frontier knee is grid-dependent and shown
only as a broadly consistent cross-check. Frontiers for $n=0.87$ and $n=1$ nearly overlap, indicating the optimum $\sigma$ is practically insensitive to $n$ over the tested range.}
\label{fig:optima}
\end{figure}

We adopt $\sigma\approx0.19$ from the normalization-independent cast-half criterion~(B).
We stress that this value is for the moderate-chroma, sRGB-typical palette used here:
$\sigma$ is a threshold in chroma units, not a universal constant, so more saturated
content or a wider gamut shifts the cast-half point upward (content-adaptive $\sigma$
scaling is left to Section~\ref{sec:future}). (A) ($\sigma\approx0.14$) is the
alternative for designs that weight hue retention more
heavily. A frontier-knee estimate is broadly consistent ($\sigma$ in $0.17$--$0.20$),
but it depends on the sweep range and axis normalization, so we do not count it as an
independent criterion. The optimum is \emph{practically $n$-insensitive}: repeating the analysis at
$n=0.87$ shifts $\sigma^{*}$ and the metrics only at the error level, which is why the
$n=1$ default (Section~\ref{sec:default}) carries no trade-off cost. The reused value
$\sigma=0.34$ (the C-axis compression fit from the author's Oklch+~\cite{oklchplus}) sits well past this moderate-palette interpolation
optimum on the cast-removal side (cast~$0.050$, gate retention~$0.30$) and is not
recommended for interpolation (Figure~\ref{fig:optima}).

\subsection{Cross-check: gate proxy validity}
\label{sec:crosscheck}
\begin{figure}[ht]
\centering
\includegraphics[width=0.82\linewidth]{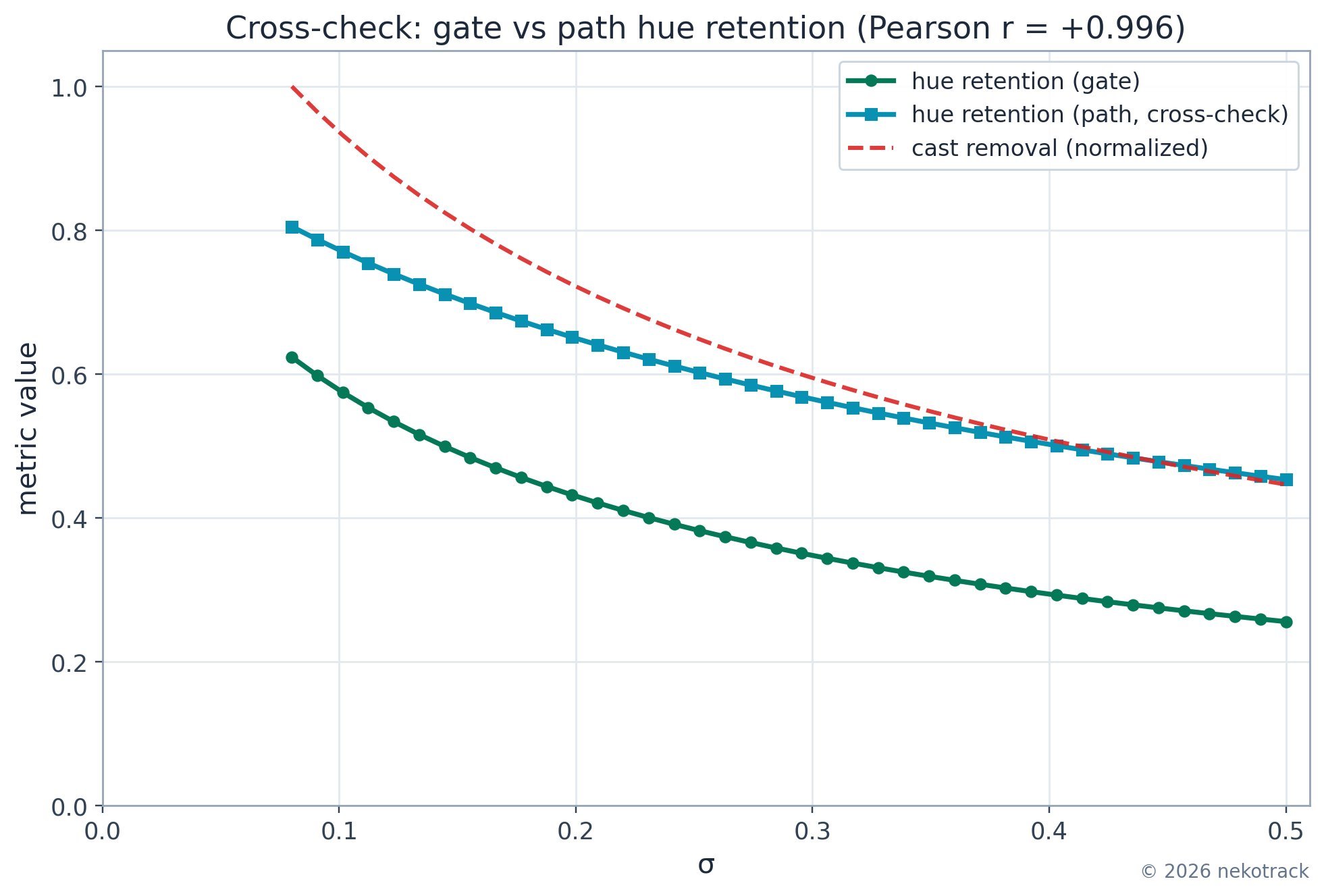}
\caption{Gate-based and path-based hue retention share the same monotone trend in
$\sigma$ and trade off against cast in the same direction, validating the gate proxy.}
\label{fig:gatepath}
\end{figure}

The hue-retention figure used above is a gate-based proxy (the mean gate $w$ over the
representative chromas $C\in\{0.10,\,0.15,\,0.20\}$). To confirm it tracks the actual interpolated paths, we also compute a
path-based hue retention---the chroma-weighted hue deviation of the COFb path from
linear Oklab, normalized by the raw-OKLCH value ($1=$ OKLCH-like, $0=$ Oklab-flat).
Across the $\sigma$ sweep the two are strongly correlated (Pearson product-moment $r=+0.996$; $r$ ranges from $-1$ to $+1$, with $+1$ a perfect linear relationship and $0$ none, so $+0.996$ means the two move almost perfectly linearly together), both
decreasing monotonically in $\sigma$, and both trade off against cast in the same
direction (cast vs.\ gate retention $+1.000$; cast vs.\ path retention $+0.998$). The
gate proxy is therefore a faithful stand-in for the path behavior, and the trade-off of
Section~\ref{sec:paramstudy} is not an artifact of the proxy (Figure~\ref{fig:gatepath}).

\section{Discussion and Limitations}
\label{sec:discussion}

\subsection{What COFb is}
COFb makes two claims, both supported in Section~\ref{sec:eval}. First, it is the only
remedy among those compared that addresses the (1) inter-hue cast: on chromatic pairs
the existing two-valued methods are numerically identical to raw OKLCH
(Table~\ref{tab:metrics}), while COFb roughly halves the path detour at the default.
Second, a single continuous, differentiable gate unifies (1) and~(2): the same
expression that treats the inter-hue detour also relaxes the achromatic-endpoint bow,
with no threshold test and no branching.

\subsection{Limitations}
We state the boundaries of the method.
\begin{itemize}
  \item \textbf{The achromatic case~(2) is better served, alone, by the existing
  switch.} NaN-inherit and the two-valued fallback drive the lateral deviation to
  exactly~0 (full linear Oklab) at an achromatic endpoint; COFb improves on raw OKLCH
  ($0.0821\to0.0313$ mean) but does not reach~0. COFb's advantage is unification across
  (1) and~(2), not superiority on~(2) in isolation. A practitioner who only needs~(2)
  and accepts a switch can keep the two-valued handling.
  \item \textbf{Chroma is not preserved.} Like any Cartesian $(a,b)$ blend, COFb grays
  slightly as the path approaches the neutral axis; the gate trades a small loss of
  saturation for removal of the cast. Methods that preserve chroma along the path would
  need a different (non-Cartesian) construction and lie outside this work's scope.
  \item \textbf{No claim of perceptual optimality.} COFb is a geometric,
  behavior-level construction defined in the $(a,b)$ plane. We do not claim its paths
  agree with a true perceptual metric, nor that $\sigma=0.19$ is perceptually optimal;
  $\sigma$ is an application-dependent parameter and we report a frontier and a
  practical default, not an optimum derived from perceptual data.
  \item \textbf{The default $\sigma$ rests on one normalization-independent criterion and
  is content-chroma dependent.}
  $\sigma\approx0.19$ is anchored on the cast-half criterion
  (Section~\ref{sec:paramstudy}) and is the value \emph{for the moderate-chroma,
  sRGB-typical palette used here}. Because $\sigma$ is a threshold in chroma units it is
  not a universal constant: more saturated content or a wider gamut moves the optimum
  upward. The frontier-knee estimate is consistent but
  grid-dependent, so we do not present it as independent corroboration. Applications
  with different priorities (e.g.\ stronger hue retention) may prefer a smaller $\sigma$
  such as the (A) operating point ($\approx0.14$). Content-adaptive $\sigma$ scaling is
  discussed in Section~\ref{sec:future}.
  \item \textbf{Endpoint $t$-derivative.} With $n=1$ the gate is $C^{1}$ in chroma up to
  the neutral axis (Section~\ref{sec:method}), which regularizes the endpoint behavior
  relative to $n<1$; we have not, however, exhaustively characterized the
  time-derivative of the rendered path at an achromatic endpoint, where $dC/dt$ and the
  gate response combine. We note this as a point for future precision rather than a
  demonstrated discontinuity.
\end{itemize}

\subsection{Relation to standards and practice}
The relevant CSS Working Group issues (\#6107 on missing hue; \#9436 / \#9224 on the
achromatic interpolation arc) are closed. We therefore position COFb as an
implementation-level reference that \emph{extends} missing-component handling
continuously rather than altering the specification. Practically, COFb is a drop-in modification of standard polar
interpolation: one gate evaluation per sample, a rational function at $n=1$
(GPU-friendly, no fractional powers, closed-form inverse), and a single parameter
$\sigma$. Because OKLCH is already the recommended interpolation space for CSS gradients
and \texttt{color-mix()} and is broadly deployed, the setting in which COFb applies is
common. Because COFb runs entirely in plain Oklab $(a,b)\!\to\!$sRGB, it needs no native
OKLCH interpolation in the target environment: it can be precomputed to sRGB stops or
evaluated in a shader, acting as a fallback that delivers the same cast-reduced gradients
where modern CSS color interpolation (or \texttt{color-mix(in oklch)}) is
unavailable---for example in older engines, image and video pipelines, or GPU shaders.

\subsection{Future work}
\label{sec:future}
Natural extensions include: a perceptual study to test whether the geometric cast
reduction corresponds to observer-judged improvement; a step-size uniformity comparison (e.g.\ the coefficient of variation of consecutive $\Delta E$), which lies on a separate axis from the geometric cast metric used here; closing the residual~(2) gap
within a continuous formulation (so that one gate also reaches the two-valued optimum at
an achromatic endpoint); and a chroma-preserving variant that removes the cast without
the near-axis graying. Because $\sigma$ is a chroma-unit threshold, a further direction
is content-adaptive $\sigma$ scaling --- a systematic hue and chroma sweep that
characterizes the distribution of the cast-half $\sigma$ and yields a gamut-independent
normalization.

\section{Conclusion}
\label{sec:conclusion}
OKLCH is the recommended space for color interpolation on the web, but its polar
parameterization casts color near the neutral axis in two ways: an inter-hue detour
between chromatic endpoints~(1) and an off-line bow at an achromatic endpoint~(2).
Existing remedies are two-valued and act only on~(2); on~(1) they reduce to raw OKLCH.
We introduced \textbf{Continuous Oklab fallback (COFb)}, a one-line chroma gate
$w(C)=C^{n}/(C^{n}+\sigma^{n})$ that continuously blends raw OKLCH toward linear Oklab
as chroma falls. A single gate reduces the (1) cast that the two-valued family leaves
untreated and unifies the handling of (1) and~(2), at the cost of a small near-axis
graying and without matching the two-valued optimum on~(2) alone. We characterized the
cast--hue trade-off, recommended a practical default ($n=1$, the rational
Michaelis--Menten form, with $\sigma\approx0.19$ from a normalization-independent
cast-half criterion), and verified the gate's properties symbolically and its proxy
metric against path-based measurement. COFb turns the existing two-valued, (2)-only
fallback into a continuous, differentiable family that additionally handles the
inter-hue case.

\appendix
\section{Reproducibility and verification}
\label{app:repro}
All symbolic claims below were verified with a computer algebra system (SymPy); all
numeric results were re-run at the paper's default $n=1,\,\sigma=0.19$. The verification
script (\texttt{gate\_symbolic\_verify.py}), the evaluation scripts, and the figure script
are available with the source.

\paragraph{Symbolic properties of the gate~\eqref{eq:gate}.}
Verified symbolically (SymPy):
\begin{align*}
&w(0^{+})=0,\qquad w(\sigma)=\tfrac12,\qquad w(\infty)=1;\\[2pt]
&w'(C)=\frac{n\,C^{\,n-1}\,\sigma^{n}}{\bigl(C^{n}+\sigma^{n}\bigr)^{2}}>0
  \quad\text{on } C>0\ \ (\text{strictly increasing});\\[2pt]
&\lim_{C\to 0^{+}} w'(C)=
  \begin{cases}\infty, & n<1,\\[1pt] 1/\sigma, & n=1,\\[1pt] 0, & n>1;\end{cases}\\[2pt]
&w\ \text{is smooth (all derivatives finite) on } C>0;\\[2pt]
&n=1:\quad w=\frac{C}{C+\sigma},\quad w'=\frac{\sigma}{(C+\sigma)^{2}},\quad
  \text{inverse } C=\frac{\sigma w}{1-w};\\[2pt]
&\text{general inverse:}\quad C=\sigma\Bigl(\tfrac{w}{1-w}\Bigr)^{1/n}.
\end{align*}

\paragraph{Reproducibility of Section~\ref{sec:eval}.}
All numeric results were re-run at the paper's default $n=1,\,\sigma=0.19$, to which the
evaluation script's default gate parameters are set, so Table~\ref{tab:metrics} is
reproduced directly. The same run reproduces the operating points of
Section~\ref{sec:paramstudy} ($\sigma=0.144$ for criterion~(A), $\sigma=0.194$ for
criterion~(B)), the weak $n$-sensitivity of the optimum (the $n=0.87$ variant reproduces
$\sigma^{*}$ and the metrics at the error level), and the cross-check correlations of
Section~\ref{sec:crosscheck} ($+0.996$, $+1.000$, $+0.998$). The frontier knee is
grid-dependent and is reported as an auxiliary cross-check only.

\begin{figure}[h]
\centering
\includegraphics[width=\linewidth]{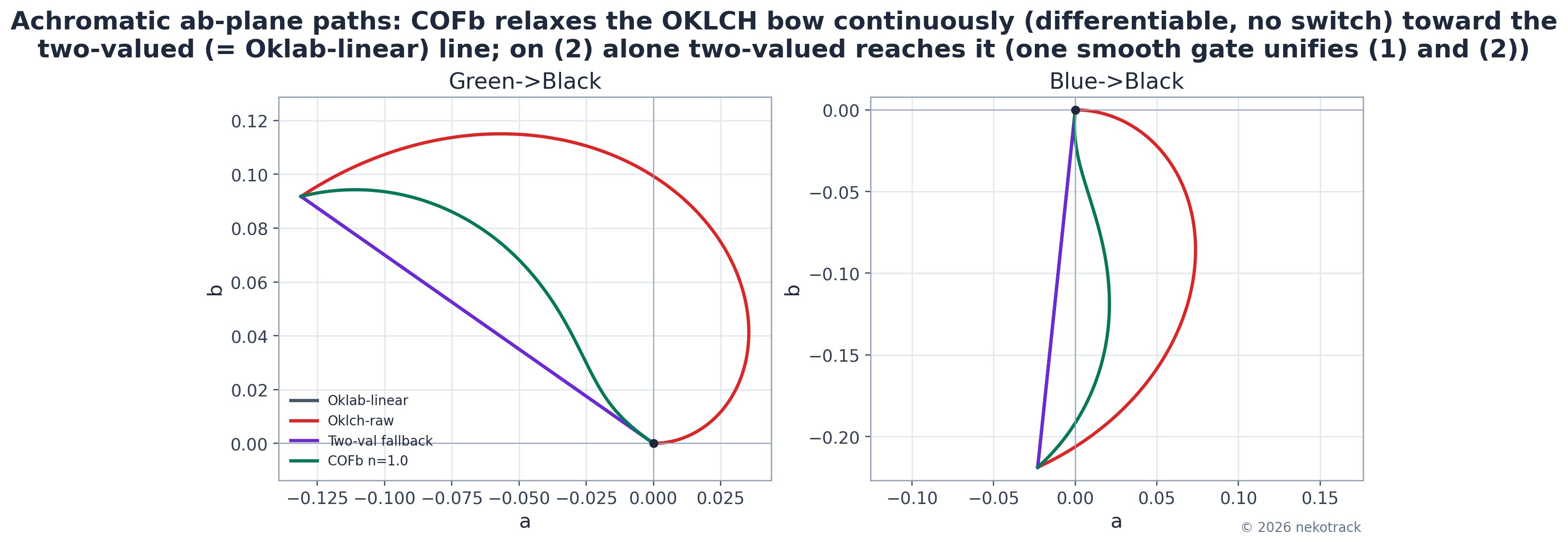}
\caption{Achromatic $(a,b)$-plane paths (supplementary). For an achromatic endpoint the
two-valued fallback equals linear Oklab and traces the straight chromatic-to-neutral
line, while raw OKLCH bows around the origin. COFb relaxes the bow continuously toward the line (differentiable, no switch) but does not reach it; on~(2) alone the two-valued switch is better, while COFb's benefit is that one smooth gate handles (1) and (2) coherently.}
\label{fig:achropaths}
\end{figure}

\paragraph{Supplementary figure: achromatic paths.}
Figure~\ref{fig:achropaths} shows the $(a,b)$ paths for the achromatic pairs,
complementing the inter-hue paths of Figure~\ref{fig:abpaths}. The two-valued fallback
coincides with linear Oklab and reaches the chromatic-to-neutral line, whereas raw OKLCH
bows around the origin (the (2) artifact); COFb relaxes the bow continuously (a differentiable gate, no switch) toward the line but does not drive it to zero, so on~(2) alone the two-valued handling is better---the advantage of COFb is that the same smooth gate handles (1) and (2) coherently (Section~\ref{sec:discussion}).

\vfill
\noindent\footnotesize Licensed under CC BY 4.0 (\url{https://creativecommons.org/licenses/by/4.0/}).

\end{document}